# Large Magnetoresistance in Co/Ni/Co Ferromagnetic Single Electron Transistors

R. S. Liu and H. Pettersson[*]

*Center for Applied Mathematics and Physics, Halmstad University, Box 823, SE-301 18 Halmstad, Sweden*

L. Michalak and C. M. Canali

*Div. of Physics, Department of Natural Sciences, Kalmar University, 391 82 Kalmar, Sweden*

D. Suyatin and L. Samuelson

*Solid State Physics/the Nanometer Structure Consortium, Lund University, Box 118, SE-22100 Lund, Sweden*

**Abstract**

We report on magnetotransport investigations of nano-scaled ferromagnetic Co/Ni/Co single electron transistors. As a result of reduced size, the devices exhibit single electron transistor characteristics at 4.2K. Magnetotransport measurements carried out at 1.8K reveal tunneling magnetoresistance (TMR) traces with negative coercive fields, which we interpret in terms of a switching mechanism driven by the shape anisotropy of the central wire-like Ni island. A large TMR of about 18% is observed within a finite source-drain bias regime. The TMR decreases rapidly with increasing bias, which we tentatively attribute to excitation of magnons in the central island.

[*] Corresponding author. E-mail: Hakan.Pettersson@ide.hh.se

The TMR observed in spintronic devices is usually defined as $TMR = \frac{R_{AP} - R_P}{R_P}$, where $R_A$ and $R_{AP}$ are the device resistances in the parallel and antiparallel magnetization configuration, respectively. TMR has been extensively investigated both experimentally and theoretically because of promising applications, e.g., magnetic random access memories (MRAM), read/write heads in hard discs as well as in various other spintronic devices[1]. More recently, a lot of research has focused on the dependence of TMR on the interplay between spin-dependent tunneling (SDT) and the Coulomb blockade effect[2] in ferromagnetic single electron transistors (F-SETs)[3]. Theoretical investigations of F-SETs predict exciting magnetoresistance properties connected to spin-accumulation on the central island[4-10]. Indeed, several experiments on F-SETs, starting with the seminal work by Ono *et al*, have demonstrated enhanced TMR, and magneto-Coulomb oscillations of the TMR as a function of external magnetic field and bias voltage[11,12]. However, due to the relatively large size of their devices, a sample temperature in the range of tens of mK was required in order to observe any significant TMR signals. For future spintronic applications it is obviously important to realize F-SETs with a considerably higher operating temperature.

In this letter, we present magnetotransport investigations of nano-scaled Co/Ni/Co F-SETs exhibiting single electron transistor characteristics at 4.2K. The fabrication yield of the device is about 50%. The devices are fabricated on top of a 100 nm thick $SiO_2$ layer, thermally grown on a Si substrate (shown in Fig. 1). The wire-like Ni islands measure 150 nm in length, 20 nm in width and 25 nm in thickness. The islands are prepared together with Ni side-gates in one step employing electron-beam lithography, followed by thermal evaporation and lift-off. Tunnel barriers of NiO are subsequently formed by $O_2$ plasma etching at a pressure of about 5 mbar for 1 minute. Ferromagnetic Co source and drain electrodes, 40 nm thick, are defined on top of the Ni islands using a high-precision alignment procedure during a second electron beam lithography step. The area of the tunnel junctions amounts to only about 40 × 20 nm². The source electrode has a length of 1.5 μm and a width of 80 nm. The corresponding dimensions of the drain electrode are 800 nm and 280 nm, respectively. The separation between the parallel drain and source electrodes is approximately 55 nm. Because of shape anisotropy, the two electrodes are expected to undergo magnetic reversal at different magnetic fields[13]. After the fabrication, conductance measurements are carried out at 4.2K in a liquid helium Dewar. The schematic circuit diagram is shown in Fig 1(c). Following these measurements, the sample is transferred to a cryostat housing a 6T superconducting magnet where magnetoresistance measurements are carried out at 1.8 K. The magnetic field is in the plane of the device with a tunable orientation with respect to the orientation of the electrodes.

Figure 2 shows a typical non-linear current-voltage (I-V) curve for the device at 4.2K. From the absence of a Coulomb staircase in the I-V curve, we conclude that the device has two symmetric tunneling junctions with the same electron tunneling rates. The modulation of the current with gate bias is shown in the inset at 4.2K and 1.8K, respectively, at a drain-source bias of 0.4 mV. Here each peak corresponds to addition of one electron to the island. The gate modulation of the current was visible up to a drain-source threshold voltage of about 2 mV at 1.8K, from which we deduce a charging energy $E_C = e^2/2C_\Sigma$ of about 1 meV.

Figure 3 shows the dependence of the resistance on the magnetic field for two sweep directions. The measurements were performed at 1.8 K with an in-plane magnetic field applied parallel to the long axis of the electrodes. From this figure two obvious interesting results are readily observed: Firstly, as the magnetic field is swept from -1.5 to 1.5 T, a substantial TMR is observed with a largest magnitude reaching ~13% between -0.15 and 0.5 T. Secondly, the TMR trace displays negative coercive fields for the two sweep directions. The origin of this phenomenon can be attributed to the special design of our device. Since the central Ni island has a wire-like shape, a substantial shape anisotropy is expected which favors a magnetic moment in the direction of the island (wire). This leads to noncollinear configurations and switching mechanisms for certain values of the external magnetic field. To explicitly clarify this point, we have performed micromagnetic modeling of the magnetization configurations of the device using the OOMMF code[14], in conjunction with rate-equations simulations of the quantum transport with noncollinear magnetization vectors of the external leads and the central island. The quantum transport simulations, shown in Fig. 3, are an extension of the usual sequential tunneling calculations for a collinear F-SET in the spirit of Ref. 8. All the relevant magnetization configurations of the device used in the transport calculations, and the values of the external magnetic field at which transitions between two configurations occur, are obtained from micromagnetic simulations. At large negative external magnetic fields, the electrodes and the central Ni island (wire) are all magnetized along the field and no TMR is observed. At -0.1T, the magnetic moment of the Ni island spontaneously relaxes to its easy magnetization axis and thus becomes perpendicular to the magnetic moments of the Co electrodes, resulting in a theoretical TMR of 8%. As the magnetic field goes through zero and subsequently increases, the two Co leads reverse their magnetization directions in a conventional manner at 0.13T (wide one) and 0.21T (narrow one), respectively, together with a progressive flip of the magnetization in the central Ni island. The theoretical value differs by a factor of ~2 from the experimentally observed TMR, which is acceptable in consideration of complex magnetization effects not included in our theoretical model. At 0.3T, the shape anisotropy of the Ni island is completely overcome by the external field and all magnetic moments are aligned parallel to the external field, resulting in minimum resistance. The magnetization configurations and corresponding resistance changes are expected to occur in an analogous manner when decreasing the magnetic field from large positive values. From Fig. 3 it is evident that there is a shift of the experimental TMR curves towards positive magnetic field, especially for the reverse sweeping curve (red one), which could be due to an exchange bias caused by the anti-ferromagnetic oxides on top of both the Ni island and the Co electrodes[15].

Fig. 4 (a) shows the current-voltage (I-V) characteristics measured at T=1.8 K at B = 0 T (black curve) and B= -1.5 T (red curve), respectively. In Fig. 4 (b) we plot the TMR, normalized by its maximum, as a function of the drain-source bias voltage (blue curve). The TMR is derived from the two I-V curves in (a) according to TMR = $\frac{I_{-1.5T} - I_{0T}}{I_{0T}}$. Interestingly, the maximum TMR reaches 18% within the Coulomb blockade regime and decreases rapidly as the drain-source bias increases above $E_c$. This abrupt decrease of TMR is not reproduced in our numerical simulations in the sequential tunneling regime, even after taking into account the decrease of spin asymmetry of the tunneling electrons for energies larger than the Fermi energy[FOOTNOTE] as shown by the purple curve of Fig. 4 (b). Such a dependence is meant to account, in a simplified way, for the bias-dependence of the density of states (DOS), which is known to influence the TMR[16,17]. Modeling microscopically this effect is very difficult, particularly since the electrodes are thermally grown which leaves no control over the crystal structure. Nevertheless, comparing our experimental results to those of Ref. 17 we note that the bias-dependence of the TMR is much stronger in our case, and we therefore hesitate to primarily attribute our results to a bias-dependent DOS. The influence of the applied electric field (bias) on the tunneling barrier height does not explain the abrupt decrease either[18, 19]. Cotunneling[5] can also be ruled out since it would result in rather a gentle decrease of the TMR with increasing bias for the relatively large ratio $k_BT/E_C \approx 0.15$ given in our case. Higher order tunneling processes can furthermore be excluded since the tunneling resistances of the investigated junctions (≈ 1 MΩ) are much larger than the quantum resistance (≈ 26 kΩ) [20].

A more likely mechanism for the decrease in TMR with increasing bias is the excitation of spin-waves or *magnons* by the spin-polarized tunneling current[21-23]. In our devices, electron spin injection orthogonal to the magnetization of the island can exert a torque acting like a transverse magnetic field which creates magnons. If many magnons are present, the island magnetization can be considerably disrupted which results in a decrease of the magnetoresistance. The excitation of magnons is an inelastic process, whose energy is provided by the bias. The dispersion of low-energy spin-waves in a ferromagnet is given by $\omega(q) = \Delta + D|\vec{q}|^2$, where $D$ is the spin-wave stiffness constant which is of the order of 500 meV in Ni [24]. The energy gap $\Delta$ is proportional to the magnetic anisotropy of the system, including shape anisotropy, and amounts to a small fraction of 1 meV. The allowed wave vectors in our small islands are discrete and can be written as $q_n = 2\pi n/L, \ n = 0,1,2...,$ where L is the size of the island. From this we estimate that some of the lowest modes can in principle be present already at small biases where the F-SET is in the Coulomb blockade regime. However, their excitation is not very efficient because the current in the off-state is small. In contrast, the current is finite in the on-state and for larger biases short wavelength magnons can be excited. A full microscopic implementation of this effect in transport is beyond the purpose of the present paper. Here we consider a phenomenological model in which the creation of spin-waves causes an exponential decrease of the spin-polarization of the island for biases above the Coulomb blockade gap. The effect of such a dependence on the normalized TMR is shown by the red curve in Fig.4(b), which seems to capture the drastic decrease of TMR with increasing bias. This scenario would also explain the persistence of TMR for biases above the Coulomb blockade gap observed in other F-SET experiments[25], where a smaller size of the grains and considerably larger tunneling resistances make the excitation of spin-waves both more energetically costly and less efficient.

In summary, we have fabricated Co/Ni/Co F-SETs exhibiting single electron transistor characteristics at 4.2 K. Magnetotransport measurements at 1.8K reveal TMR traces with negative coercive fields which we interpret in terms of a switching mechanism driven by shape anisotropy in the central wire-like Ni island. A large TMR signal of a maximum magnitude of 18% has been observed within a small drain-source bias regime. The TMR decreases rapidly with increasing bias, which we tentatively attribute to an excitation of magnons in the central island.

The authors thank Arne Brataas and Allan MacDonald for fruitful discussions. The authors furthermore acknowledge financial support from Halmstad University, the Faculty of Natural Sciences at Kalmar University, the Swedish Research Council under Grant No: 621-2004-4439, the Swedish National Board for Industrial and Technological Development, the Office of Naval Research, and the Swedish Foundation for Strategic Research.

**References:**

[1] *Spin Electronics,* edited by M. Ziese and M.J. Thornton (Springer, Berlin Heidelberg, 2001).

[2] *Single Charge Tunneling*, NATO ASI Series, Vol. 294, edited by H. Grabert and M. H. Devoret (Plenum, New York, 1992).

[3] *Spin Dependent Transport in Magnetic Nanostructures*, edited by S. Maekawa and T. Shinjo (Taylor and Francis, London, 2002).

[4] J. Barnas and A. Fert, Phys. Rev. Lett. **80**, 1058 (1998).

[5] S. Takahashi and S. Maekawa, Phys. Rev. Lett. **80**, 1758 (1998).

[6] J. Barnas and A. Fert, Europhys. Lett. **44**, 85 (1998).

[7] A. Brataas, Y. V. Nazarov, J. Inoue and G. E. W. Bauer, Phys. Rev. B **59**, 93(1999).

[8] A. N. Korotkov and V. I. Safarov, Phys. Rev. B **59**, 89 (1999).

[9] H. Imamura, S. Takahashi and S. Maekawa, Phys. Rev. B **59**, 6017(1999).

[10] I. Weymann and J. Barnas, Phys. Stat. Solidi B **236**, 561 (2003).

[11] K. Ono, H. Shimada, and Y. Ootuka, J. Phys. Soc. Jpn. **66**, 1261 (1997).

[12] K. Ono, H. Shimada, and Y. Ootuka, J. Phys. Soc. Jpn. **67**, 2852 (1998).

[13] S.Y. Chou, P. R. Krauss, and L. Kong, J. Appl. Phys. **79**, 6101(1996).

[14] OOMMF is Object Oriented Micromagnetic Framework, a micromagnetic simulation code available free from NIST at http://math.nist.gov/oommf/.

[15] For a review on exchange bias see: J Nogues and I K Schuller, J. Magn. Magn. Mater. **192**, 203 (1999); A. E. Berkowitz and K Takano J. Magn. Magn. Mater. **200**, 552 (1999).

[FOOTNOTE] In the simulations we assume that the spin polarization of both external electrodes and central island decays exponentially for energies larger than their corresponding Fermi energy in units of $E_c$.

[16] P. LeClair, J. T. Kohlhepp, C. H. van de Vin, H. Wieldraaijer, H. J. M. Swagten, W. J. M. de Jonge, A. H. Davis, J. M. MacLaren, J. S. Moodera, and R. Jansen, Phys. Rev. Lett. **88**, 107201 (2002).

[17] X. H. Xiang, T. Zhu, J. Du, G. Landry and J. Q. Xiao, Phys. Rev. B **66**, 174407 (2002).

[18] X. Zhang, B.-Z. Li, G. Sun and F.-C. Pu, Phys. Rev. B **56**, 5484 (1997).

[19] A. M. Bratkovsky, Phys. Rev. B **56**, 2344 (1997).

[20] X. H. Wang and A. Brataas, Phys. Rev. Lett. **83**, 5138 (1999).

[21] L. Berger, Phys. Rev. B **54**, 9353 (1996).

[22] J. C. Slonczewski, J. Mag. Magn. Mat **195,** L1 (1996),

[23] S.I Kiselv, J.C. Sankey, I.N. Krivorotov, N.C. Emley, A.G.F. Garcia, R.A. Burhman and D.C. Ralph, Phys. Rev. B **72** 064430 (2005).

[24] J. Kubler, *The theory of Itinerant electron magnetism,* Oxford Science, 2000.

[25] K. Yakushiji, F. Ernult, H. Imamura, K. Yamane, S. Mitani, K. Takanashi, S. Takahashi, S. Maekawa and H. Fujimori, Nat. Mater. **4**, 57 (2005).

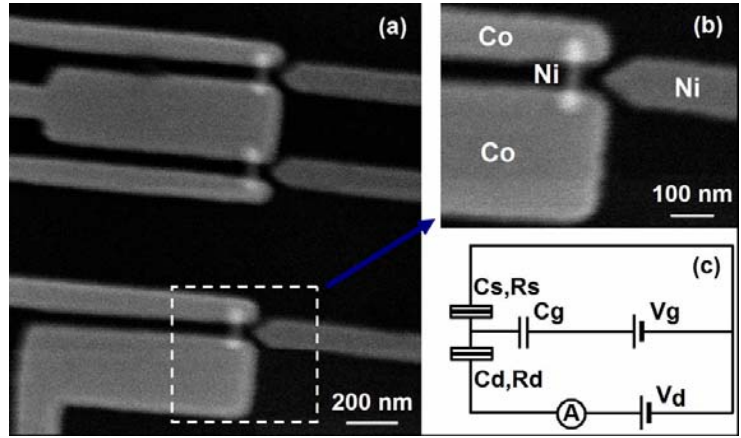

Figure 1. (a) Top-view SEM image of the device geometry. (b) Expanded view of an isolated F-SET marked by a white short-dash rectangle in (a). The Ni island has dimensions of 150 nm (length) × 20 nm (width) × 25 nm (thickness). (c) Circuit diagram of the connected device.

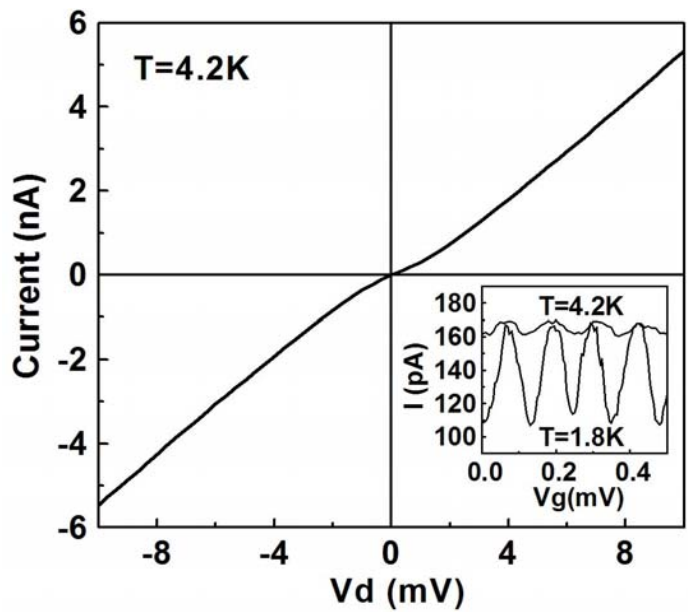

Figure 2. Nonlinear current-voltage (I-V) characteristics at 4.2K recorded at zero magnetic field with gate voltage $V_g$ = 0 V. Lower right inset: Gate modulated current curves at 4.2 K (upper) and 1.8K (lower), respectively, at an applied drain-source bias of 0.4 mV.

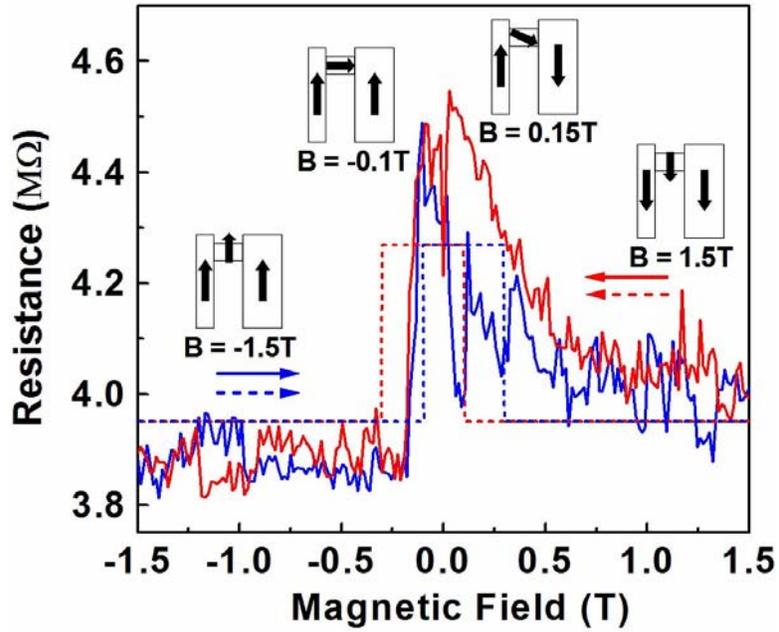

Figure 3. The dependence of the resistance on the magnetic field at 1.8K with an applied drain-source bias of 1.5 mV. The solid and dotted curves are experimental data and simulation results, respectively. The magnetic field sweep directions are denoted with horizontal arrows in different colors. The insets show schematic magnetization configurations of the device obtained from micromagnetic modeling, sweeping the magnetic field from -1.5T to +1.5T.

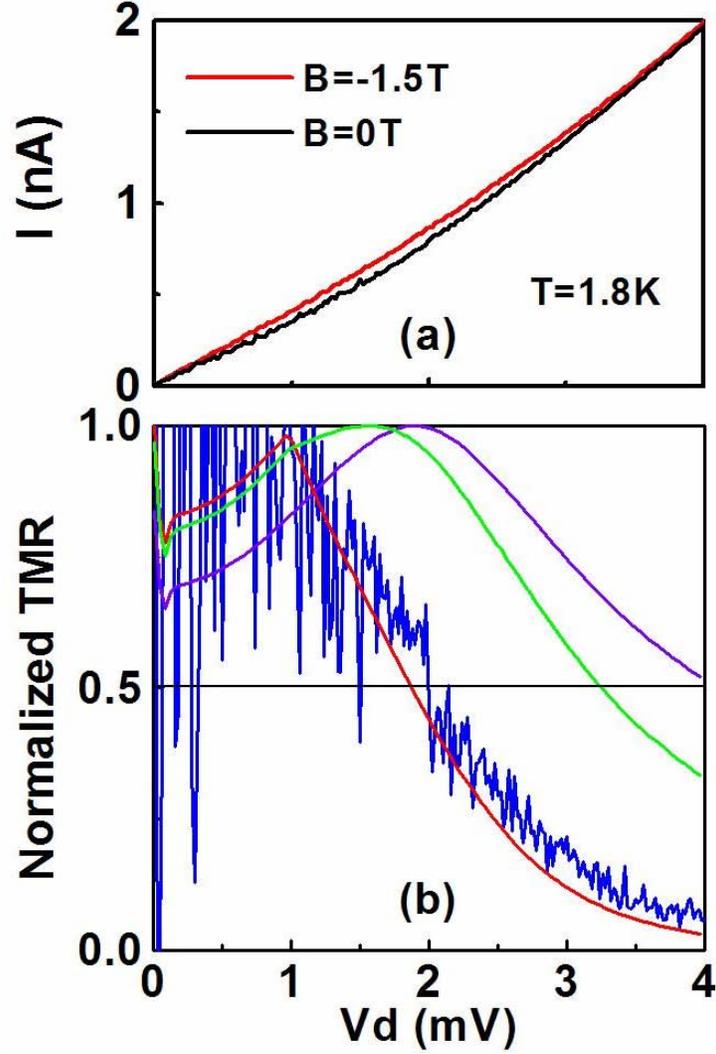

Figure 4. (a) Current-voltage characteristics measured at 1.8K at B = 0 T (black curve) and B= -1.5 T (red curve), respectively. (b) The dependence of TMR = ($I_{B=-1.5T}$ - $I_{B=0T}$)/$I_{B=0T}$, normalized by its maximum value, on drain-source voltage. The current values are taken from the two I-V curves in (a). The red, green and purple solid curves are the results of numerical simulations in which the spin polarization, $P(V)$, of the central island is assumed to decay exponentially for biases larger than the Coulomb blockade gap according to the expression $P(V) = P_0 e^{[(E_c - eV)/\gamma E_c]}$. The three curves correspond to $\gamma = 1, 5, \infty$ (no decay), respectively.